\documentclass[a4paper, 12pt]{article}%
\usepackage{amsmath}
\usepackage{graphicx}%
\usepackage{amsfonts}%
\usepackage{amssymb}

\begin{document}
\noindent{\Huge Evolution integrals}

\bigskip

\noindent\textbf{Rocco Duvenhage}

\bigskip

\noindent Department of Mathematics and Applied Mathematics

\noindent University of Pretoria, 0002 Pretoria, South Africa.

\noindent E-mail: rocco@postino.up.ac.za

\bigskip

\noindent6 April 2006

\bigskip

\noindent\textbf{Abstract: }A framework analogous to path integrals in quantum
physics is set up for abstract dynamical systems in a $W^{\ast}$-algebraic
setting. We consider spaces of evolutions, defined in a specific way, of a
$W^{\ast}$-algebra $A$ as an analogue of spaces of classical paths, and show
how integrals over such spaces, which we call ``evolution integrals'', lead to
dynamics in a Hilbert space on a ``higher level'' which is viewed as an
analogue of quantum dynamics obtained from path integrals. The measures with
respect to which these integrals are performed are projection valued.

\section{Introduction}

Path integrals in quantum physics essentially express the dynamics of a
quantum system as an integral over a space of paths in the configuration space
(or in the phase space) of the corresponding classical system; see for example
[PS], or many other standard texts on quantum physics. Also see [F] for the
early work on this topic. In this paper the goal is to set up an analogous
framework for abstract dynamical systems, where the dynamics of a system on a
``higher level'' is expressed in terms of an integral over a space of
evolutions of a system on a ``lower level''. The ``lower level'' will be given
by a $W^{\ast}$-algebra while the ``higher level'' will be expressed in terms
of a Hilbert space. We will call such integrals ``evolution integrals'', and
they will be defined in Section 4.

By an abstract dynamical system we mean a pair $(A,\alpha)$ where $A$ is a
$C^{\ast}$-algebra and $\alpha$ is a representation $G\rightarrow
$Aut$(A):g\mapsto\alpha_{g}$ of some group $G$ in the automorphism group
Aut$(A)$ of $A$. We can refer to $\alpha$ as the \textit{evolution} of the
system. But for a given $G$, we can have different evolutions of $A$, namely
different representations of $G$ in Aut$(A)$.

All of these evolutions by definition have group properties. However, in path
integrals not all the paths in the space of paths over which we integrate, can
be expected to be a segment from a possibly longer path which has group
properties. For example a path may intersect itself. Therefore we will also
allow more general evolutions in the case of abstract dynamical systems, where
the group structure doesn't play a role anymore. First write $\alpha$ as
$(\alpha_{g})_{g\in G}$. Note that at every point in ``time'', $g\in G$, an
element $a\in A$ at ``time'' $e\in G$ (the identity element) will have evolved
to $\alpha_{g}(a)$ by means of the $\ast$-automorphism $\alpha_{g}$. If we
want to retain this fact, but ignore the group structure, we can generalize
evolutions by allowing $(\alpha_{g})_{g\in G}$ to be viewed as an evolution
even when $G\ni g\mapsto\alpha_{g}\in$Aut$(A)$ is an arbitrary function. This
is a very wide class of functions, but because of the methods we employ in
this paper (see Section 3) we will not restrict ourselves to smaller classes,
for example functions which are continuous in some specified topologies on $G$
and Aut$(A)$. Of course, we are now no longer working with group
representations, and in fact we will not make use of the group structure of
$G$ or Aut$(A)$ at all in this paper. Therefore we will replace $G$ by an
arbitrary set $T$. In Section 4 we will describe how a grouplike structure
emerges on the ``higher level'' when we consider certain collections of sets
$T$ in a measure space $(\mathfrak{T},\Sigma,\mu)$.

We will in fact have to generalize even further, again because of the methods
we use, and replace Aut$(A)$ by its closure $\overline{\text{Aut}(A)}$, which
will be compact, in an appropriate topology on the space of bounded linear
operators on $A$. In order to do be able to do this, we will take $A$ to be
$W^{\ast}$-algebra, rather than just a $C^{\ast}$-algebra. The evolutions of
$A$ that we will consider, will therefore be all functions of the form $T\ni
t\mapsto\alpha_{t}\in\overline{\text{Aut}(A)}$. We discuss this space of
evolutions, namely $\overline{\text{Aut}(A)}^{T}$, in more detail in Section 2.

As will be explained in Section 4, the ``evolution integrals'' over
$\overline{\text{Aut}(A)}^{T}$ will differ from path integrals in an important
respect. In the former case we will use a projection valued measure obtained
by spectral theory and the representation theory of $C^{\ast}$-algebras
applied to $C\left(  \overline{\text{Aut}(A)}^{T}\right)  $, which we discuss
in Section 3, but this measure will not be unique. We will in fact have many
different measures, all of them natural (or canonical) in a sense to be
explained, and all of them being unitary transformations of one another.
Another difference is that the system we start off with (the ``lower level'')
need not abelian, that is to say classical. In other words $A$ can be a
noncommutative $W^{\ast}$-algebra. The analogy between path integrals and
integrals over $\overline{\text{Aut}(A)}^{T}$ is clarified in Section 5.

The mathematical results of this paper are concentrated in Sections 2 and 3
where we develop the mathematical tools that will enable us to set up the
basic framework of evolution integrals in Section 4. Sections 4 and 5 consist
out of definitions, discussion and some simple results describing evolution
integrals and their meaning. We will not look at any applications of evolution
integrals in this paper.

\section{The evolution space}

In this section, $A$ will denote an arbitrary $W^{\ast}$-algebra, in other
words a $C^{\ast}$-algebra which when viewed as a Banach space has some Banach
space as a predual; see for example [S]. It is known that this predual of $A$
is unique, and we will denote it by $A_{\ast}$. Let Aut$(A)$ denote the set of
all $\ast$-automorphisms of $A$.

We will use the following notation: For any normed spaces $X$ and $Y$, let
$B(X\times Y)$ denote the Banach space of all bounded bilinear mappings
$X\times Y\rightarrow\mathbb{C}$, and $L(X,Y)$ the normed space of all bounded
linear mappings $X\rightarrow Y$. Furthermore we set $L(X):=L(X,X)$. A unit
ball will be indicated by $X_{1}:=\left\{  x\in X:\left\|  x\right\|
\leq1\right\}  $. When $X$ and $Y$ are Banach spaces, we will denote their
projective tensor product (see for example [R]) by $X\hat{\otimes}_{\pi}Y$.

\bigskip

\noindent\textbf{Theorem 2.1. }\textit{The space }$L(A)$ \textit{has a Banach
space predual }$A\hat{\otimes}_{\pi}A_{\ast}$\textit{, which gives a weak*
topology on }$L(A)$\textit{ in which the closure }$\overline{\text{Aut}(A)}%
$\textit{ of }Aut$(A)$\textit{ is a compact Hausdorff space.}

\bigskip

\noindent\textit{Proof.} We have canonical isometric isomorphisms (which we
can also call Banach space isomorphisms)
\begin{equation}
\left(  A\hat{\otimes}_{\pi}A_{\ast}\right)  ^{\ast}\cong B(A\times A_{\ast
})\cong L\left(  A,(A_{\ast})^{\ast}\right)  =L(A) \tag{2.1}%
\end{equation}
so $L(A)$ has $A\hat{\otimes}_{\pi}A_{\ast}$ as a predual. By Alaoglu's
theorem the unit ball $L(A)_{1}$ is compact in the weak* topology thus
obtained on $L(A)$, and of course the weak* topology on $L(A)$ is Hausdorff,
since a predual separates the points of a space. In particular then,
$L(A)_{1}$ is weak* closed in $L(A)$. By definition any $\alpha\in$Aut$(A)$ is
a linear mapping $A\rightarrow A$, but since it is a $\ast$-isomorphism and
$A$ a $C^{\ast}$-algebra, we also know that it is norm preserving, so
$\left\|  \alpha\right\|  =1$. Hence Aut$(A)\subset L(A)_{1}$. Since
$L(A)_{1}$ is weak* closed, we have $\overline{\text{Aut}(A)}\subset L(A)_{1}$
for the weak* closure of Aut$(A)$ in $L(A)$, and therefore $\overline
{\text{Aut}(A)}$ is compact and Hausdorff. $\square$

\bigskip

By Tychonoff's theorem we then immediately have:

\bigskip

\noindent\textbf{Corollary 2.2.} \textit{For any set }$T$\textit{, the product
space }$\overline{\text{Aut}(A)}^{T}:=\prod_{t\in T}\overline{\text{Aut}(A)}%
$\textit{ is compact and Hausdorff.}

\bigskip

Because $A_{\ast}$ is the unique predual of $A$, the predual $A\hat{\otimes
}_{\pi}A_{\ast}$ is canonical in the sense that the Banach space isomorphisms
in (2.1) are canonical. In this sense we can view the weak* topology in which
the closure $\overline{\text{Aut}(A)}$ was taken, as a natural weak* topology
on $L(A)$. It is however unclear whether $A\hat{\otimes}_{\pi}A_{\ast}$ is the
unique predual of $L(A)$; for example this type of problem has been studied in
[GS], but assuming the Radon-Nikod\'{y}m property for $A$ and $A_{\ast}$,
which unfortunately we do not have in general [CI], [C].

We will call $\overline{\text{Aut}(A)}^{T}$ in Corollary 2.2, the
\textit{evolution space} of $A$ over the set $T$.

Theorem 2.1 is the only place in this paper where we use the properties of
$\ast$-automorphisms of $A$. From now on we will only use the fact that
$\overline{\text{Aut}(A)}^{T}$ is a compact Hausdorff space.

\section{Natural spectral measures}

In path integrals the measure can be heuristically viewed as an infinite
dimensional Lebesgue measure. In particular we want such a measure be natural
in a certain sense, and intuitively we want it to assign the same weight to
each path. In the case of an abstract dynamical system we would similarly like
to obtain measures on $\overline{\text{Aut}(A)}^{T}$ which are in some way
natural and intuitively assign the same weight to each evolution. We will
approach this problem using the representation theory of $C^{\ast}$-algebras,
in particular for the $C^{\ast}$-algebra $C(K)$ of continuous functions
$K\rightarrow\mathbb{C}$ with $K$ a compact Hausdorff space. We will in fact
obtain projection valued measures, and this will be done using the following
result, which is the reason why compact Hausdorff spaces play an important
role in this paper:

If $K$ is a compact Hausdorff space, $H$ a Hilbert space, and $\varphi
:C(K)\rightarrow L(H)$ a unital $\ast$-homomorphism, then there is a unique
spectral measure $E$ relative to $(K,H)$ such that
\begin{equation}
\varphi(f)=\int_{K}fdE \tag{3.1}%
\end{equation}
for all $f\in C(K)$, where by a \textit{spectral measure} relative to $(K,H)$
we mean a map $E$ from the $\sigma$-algebra of Borel sets of $K$ to the set of
projections in $L(H)$ such that $E(\varnothing)=0$, $E(K)=1$, $E(V_{1}\cap
V_{2})=E(V_{1})E(V_{2})$ for all Borel $V_{1},V_{2}\subset K$, and for all
$x,y\in H$ the function $E_{x,y}:V\mapsto\left\langle x,E(V)y\right\rangle $
is a regular complex Borel measure on $K$; see for example [M]. We will refer
to $E$ as the \textit{spectral resolution} of $\varphi$, and will sometimes
denote it by $E_{\varphi}$. Note that the integral $\int_{K}fdE$ is defined
for all bounded complex-valued Borel functions $f$ on $K$, the space of such
functions being denoted by $B_{\infty}(K)$ which with the sup-norm is a
$C^{\ast}$-algebra, by demanding that $\left\langle x,\left(  \int
_{K}fdE\right)  y\right\rangle =\int_{K}fdE_{x,y}$ for all $x,y\in H$. Besides
being well defined, this integral in fact also allows us to use (3.1) to
naturally extend $\varphi$ to a unital $\ast$-homomorphism $\tilde{\varphi
}:B_{\infty}(K)\rightarrow L(H):f\mapsto\int_{K}fdE$. We will from now on
consistently use this notation to denote the extension $B_{\infty
}(K)\rightarrow L(H)$ of a unital $\ast$-homomorphism $C(K)\rightarrow L(H)$
given by its spectral resolution.

However, we will require a $\varphi$ which is in some way a canonical
representation of $C(K)$. Since $C(K)$ is an abelian $C^{\ast}$-algebra, and
$K$ is a compact Hausdorff space, the set of pure states of $C(K)$ can be
identified with $K$ via
\[
\omega_{x}(f):=f(x)
\]
which defines a pure state $\omega_{x}$ on $C(K)$ corresponding to $x\in K$.
Let $(H_{x},\pi_{x},\Omega_{x})$ be the GNS representation of $C(K)$
associated to $\omega_{x}$, namely $H_{x}$ is a Hilbert space (which in this
case happens to be one dimensional), $\pi_{x}:C(K)\rightarrow L(H_{x})$ is a
$\ast$-homomorphism, and $\Omega_{x}\in H_{x}$ is a cyclic vector for $\pi
_{x}$ such that $\omega_{x}(f)=\left\langle \Omega_{x},\pi_{x}(f)\Omega
_{x}\right\rangle $ for all $f\in C(K)$. Now consider the direct sum of all
such representations, namely%
\[
H:=\bigoplus_{x\in K}H_{x}%
\]
and
\[
\pi:=\bigoplus_{x\in K}\pi_{x}%
\]
which we will call a \textit{pure representation} $(H,\pi)$ of $C(K)$. This of
course is a faithful representation by the standard representation theory of
$C^{\ast}$-algebras (see for example [KR]), and it is straightforward to see
that it is unital, i.e. $\pi(1)=1\in L(H)$. We can view this representation as
being canonical, since no pure state is given preference over another. The
spectral resolution $E_{\pi}$ can in this sense then be viewed as a natural
spectral measure defined on the Borel $\sigma$-algebra of $K$. Note however
that such a pure representation is not quite unique, since the GNS
representation is only unique up to unitary equivalence, where this unitary
operator can be from one Hilbert space to another. For example, if
$U:H\rightarrow H$ is unitary, then $U^{\ast}\pi(\cdot)U$ is also a pure
representation of $C(K)$ obtained out of unitary transformations of the GNS
representations above from the $H_{x}$'s to other subspaces of $H$. Hence the
spectral measure that we obtain is also not unique, despite being natural.
This lack of uniqueness has a role to play, as we will briefly discuss in
Section 4.

The unit vector $\Omega_{x}$ can also be viewed as an element of $H$ in the
obvious canonical way: $\Omega_{x}^{\prime}:=\left(  \Omega_{x,y}^{\prime
}\right)  _{y\in K}$ with $\Omega_{x,y}^{\prime}=\Omega_{x}$ when $y=x$, and
$\Omega_{x,y}^{\prime}=0$ otherwise. Hence $\Omega_{x}^{\prime}$ represents
$\omega_{x}$ as a vector in $H$, namely $\omega_{x}(f)=\left\langle \Omega
_{x},\pi_{x}(f)\Omega_{x}\right\rangle =\left\langle \Omega_{x}^{\prime}%
,\pi(f)\Omega_{x}^{\prime}\right\rangle $ for all $f\in C(K)$. The vectors
$\Omega_{x}^{\prime}$, $x\in K$, are orthonormal in $H$. Furthermore, since
$C(K)$ is an abelian $C^{\ast}$-algebra and $\omega_{x}$ is a pure state, it
follows that $H_{x}$ is one dimensional for every $x\in K$. This means that
$\left\{  \Omega_{x}^{\prime}\right\}  _{x\in K}$ is a total orthonormal set
in $H$, i.e. it is an orthonormal basis for $H$. This gives us a simple
interpretation for $H$, namely it has an orthonormal basis $\left\{
\Omega_{x}^{\prime}\right\}  _{x\in K}$ whose elements represent the points of
$K$.

We now want to argue that $E:=E_{\pi}$ attaches the same ``weight'' to each of
$K$ 's elements, and more generally that if the Borel sets $V_{1},V_{2}\subset
K$ are in some intuitive sense ``equally big'', then the projections
$E(V_{1})$ and $E(V_{2})$ are ``equally big''. We do this via the following
theorem, in which span$M$ denotes the space of finite linear combinations of
elements of the set $M$ in a vector space, and $\overline{\text{span}M}$ its
norm closure in case of a normed space:

\bigskip

\noindent\textbf{Theorem 3.1. }\textit{Let }$\left(  H,\pi\right)  $\textit{
be a pure representation of }$C(K)$\textit{ where }$K$\textit{ is a compact
Hausdorff space, and let }$E$\textit{ be the spectral resolution of }$\pi
$\textit{. Let }$(H_{x},\pi_{x},\Omega_{x})$\textit{ for }$x\in K$\textit{ be
the GNS representations of which }$\left(  H,\pi\right)  $\textit{ is the
direct sum (as above). Represent each }$\Omega_{x}$\textit{ in the canonical
way as an element of }$H$\textit{, and still denote it by }$\Omega_{x}%
$\textit{. It then follows that }$E(V)$\textit{ is the projection of }%
$H$\textit{ onto the subspace }%
\[
\overline{\text{span}\left\{  \Omega_{x}:x\in V\right\}  }%
\]
\textit{for any Borel }$V\subset K$\textit{.}

\bigskip

\noindent\textit{Proof.} Consider any $x\in K$ and define the positive Borel
measure $E_{x}$ on $K$ by $E_{x}(V)=\left\langle \Omega_{x},E(V)\Omega
_{x}\right\rangle $. The first thing to notice is that from $E(K)=1$ and
$\left\|  \Omega_{x}\right\|  =1$ we have $E_{x}(K)=1$. Since $E_{x}$ is
regular as mentioned above, we can in particular approximate $E_{x}\left(
\{x\}\right)  $ from above, namely for any $\varepsilon>0$ there exists an
open set $V_{0}\subset K$ containing $x$ such that
\begin{equation}
E_{x}(V_{0})<E_{x}\left(  \{x\}\right)  +\varepsilon\text{.} \tag{3.2}%
\end{equation}
Since $\{x\}$ is closed and $K$ is normal, we know by Urysohn's lemma that
there is a continuous $f:K\rightarrow\lbrack0,1]$ such that $f(x)=1$ and
$f|_{K\backslash V_{0}}=0$. Furthermore, since $f\in C(K)$, it follows that
\begin{align*}
\int_{K}fdE_{x}  &  =\left\langle \Omega_{x},\pi(f)\Omega_{x}\right\rangle \\
&  =f(x)\text{.}%
\end{align*}
Also note from $\chi_{\{x\}}\leq f\leq\chi_{V_{0}}$, where $\chi$ denotes
characteristic functions, that $\int_{K}\chi_{\{x\}}dE_{x}\leq\int_{K}%
fdE_{x}\leq\int_{K}\chi_{V_{0}}dE_{x}$. Hence $E_{x}\left(  \{x\}\right)  \leq
f(x)\leq E_{x}\left(  V_{0}\right)  $, which when combined with (3.2) gives
\[
f(x)-\varepsilon<E_{x}\left(  \{x\}\right)  \leq f(x)
\]
and since $\varepsilon>0$ was arbitrary, this means that $E_{x}\left(
\{x\}\right)  =f(x)=1$. Thus for any Borel $V\subset K$ containing $x$ we have
$1=E_{x}\left(  \{x\}\right)  \leq E_{x}\left(  V\right)  \leq E_{x}(K)=1$, so
$\left\|  E(V)\Omega_{x}\right\|  ^{2}=E_{x}(V)=1=\left\|  \Omega_{x}\right\|
^{2}$ hence $E(V)\Omega_{x}=\Omega_{x}$, since $E(V)$ is a projection. For any
Borel $V\subset K$ not containing $x$, on the other hand, it follows that
$\left\|  E(V)\Omega_{x}\right\|  ^{2}=E_{x}(V)=1-E_{x}(K\backslash V)=0$, so
$E(V)\Omega_{x}=0$. We have therefore shown that
\[
E(V)\Omega_{x}=\left\{
\begin{array}
[c]{l}%
\Omega_{x}\text{ if }x\in V\\
0\text{ if }x\in K\backslash V
\end{array}
\right.
\]
for all Borel $V\subset K$. However, we have already argued that the vectors
$\Omega_{x}$, $x\in K$, form a total orthonormal set in $H$, which completes
the proof. $\blacksquare$

\bigskip

Now we argue intuitively as follows: By Theorem 3.1, $E\left(  \{x\}\right)  $
is the projection onto $\mathbb{C}\Omega_{x}$ for every $x\in K$, hence it
seems clear that $E$ attaches an equal ``weight'' to every point of $K$,
namely projections of equal size, in this case one-dimensional. More
generally, if the Borel sets $V_{1},V_{2}\subset K$ are ``equally big'', by
which we intuitively mean they have ``the same number of points'', then
Theorem 3.1 tells us that $E(V_{1})H$ and $E(V_{2})H$ are ``equally big'' in
the sense that they are both spanned by ``the same number of vectors'' from
the total orthonormal set $\left\{  \Omega_{x}:x\in K\right\}  $, i.e. we can
view the projections $E(V_{1})$ and $E(V_{2})$ as being equally big.

We will apply these ideas in the case where $K$ is a product space of the form
$\overline{\text{Aut}(A)}^{T}$. Therefore we next study spectral measures as
obtained above, in the case of topological product spaces.

\bigskip

\noindent\textbf{Theorem 3.2.} \textit{Let }$\mathfrak{T}$\textit{ be any
non-empty set, and }$K_{t}$\textit{ a compact Hausdorff space for every }%
$t\in\mathfrak{T}$\textit{. Set }%
\[
K_{T}:=\prod_{t\in T}K_{t}%
\]
\textit{with the product topology for every }$T\subset\mathfrak{T}$\textit{,
write }$K:=K_{\mathfrak{T}}$\textit{, let }$\iota_{T}:K\rightarrow
K_{T}:\left(  x_{t}\right)  _{t\in\mathfrak{T}}\mapsto\left(  x_{t}\right)
_{t\in T}$\textit{ and define }$\psi_{T}:C(K_{T})\rightarrow C(K):f\mapsto
f\circ\iota_{T}$\textit{. Let }$H$\textit{ be any Hilbert space and }%
$\varphi:C(K)\rightarrow L(H)$\textit{ any unital }$\ast$%
\textit{-homomorphism, and set }$\varphi_{T}:=\varphi\circ\psi_{T}$\textit{
for every }$T\subset\mathfrak{T}$\textit{. Let }$E_{T}$\textit{ be the
spectral resolution of }$\varphi_{T}$\textit{ and write }$E:=E_{\mathfrak{T}}%
$\textit{, i.e. }$E$\textit{ is the spectral resolution of }$\varphi$\textit{.
Then we have }%
\[
E_{T}=E\circ\iota_{T}^{-1}%
\]
\textit{for all }$T\subset\mathfrak{T}$\textit{ with }$T\neq\varnothing$\textit{.}

\bigskip

\noindent\textit{Proof.} The case $T=\mathfrak{T}$ is trivial, so we will
assume $T\neq\mathfrak{T}$. This will ensure that products written as
$K_{T}\times K_{\mathfrak{T}\backslash T}$ are not trivial.

For any $x\in H$ we know from previous remarks that $E_{T,x}:=\left(
E_{T}\right)  _{x,x}$ and $E_{x}:=E_{x,x}$ are regular positive Borel
measures. Since $\iota_{T}$ is continuous, we can similarly define
\begin{align*}
\left(  E\circ\iota_{T}^{-1}\right)  _{x}(V)  &  :=\left\langle x,\left(
E\circ\iota_{T}^{-1}\right)  (V)x\right\rangle \\
&  =\left\langle x,E\left(  \iota_{T}^{-1}(V)\right)  x\right\rangle \\
&  =E_{x}\circ\iota_{T}^{-1}(V)
\end{align*}
for every Borel $V\subset K_{T}$, from which it is clear that $F_{T,x}%
:=\left(  E\circ\iota_{T}^{-1}\right)  _{x}=E_{x}\circ\iota_{T}^{-1}$ is a
positive Borel measure on $K_{T}$. Note that here we use the notation
$F_{T}=E\circ\iota_{T}^{-1}$. We now firstly prove that $F_{T,x}$ is regular:

Consider any Borel $V\subset K_{T}$, then $\iota_{T}^{-1}(V)$ is a Borel set
in $K$, but $E_{x}$ is regular, so for any $\varepsilon>0$ there exists a
compact set $V_{1}\subset K$ and an open set $V_{0}\subset K$ such that
$V_{1}\subset\iota_{T}^{-1}(V)\subset V_{0}$ and
\[
E_{x}(V_{0})-\varepsilon<E_{x}\left(  \iota_{T}^{-1}(V)\right)  <E_{x}%
(V_{1})+\varepsilon\text{.}%
\]
Since $\iota_{T}$ is continuous, $V_{1}^{\prime}:=\iota_{T}(V_{1})$ is
compact, and clearly we also have $V_{1}^{\prime}\subset\iota_{T}\left(
\iota_{T}^{-1}(V)\right)  \subset V$ (in fact, the last inclusion is equality,
since $\iota_{T}$ is a surjection) and $V_{1}\subset\iota_{T}^{-1}%
(V_{1}^{\prime})$ so $E_{x}(V_{1})\leq F_{T,x}(V_{1}^{\prime})$. With the set
$V_{0}$ we need to be a bit more careful. We need an open $V_{0}^{\prime
}\subset K_{T}$ such that $V\subset V_{0}^{\prime}$ and $\iota_{T}^{-1}%
(V_{0}^{\prime})\subset V_{0}$. Since $\iota_{T}$ is a projection, we have
$\iota_{T}^{-1}(V)=V\times K_{\mathfrak{T}\backslash T}$ while
$K_{\mathfrak{T}\backslash T}$ is compact by Tychonoff's theorem. Hence, for
every $v\in V$ we have $\{v\}\times K_{\mathfrak{T}\backslash T}\subset V_{0}%
$, and the tube lemma says that there is a ``tube'' $N_{v}\times
K_{\mathfrak{T}\backslash T}$ with $N_{v}$ an open neighbourhood of $v$ in
$K_{T}$ such that $\{v\}\times K_{\mathfrak{T}\backslash T}\subset N_{v}\times
K_{\mathfrak{T}\backslash T}\subset V_{0}$. Let $V_{0}^{\prime}:=\bigcup_{v\in
V}N_{v}$, which is then an open set in $K_{T}$ such that $V=\bigcup_{v\in
V}\{v\}\subset V_{0}^{\prime}$, and $\iota_{T}^{-1}(V_{0}^{\prime}%
)=\bigcup_{v\in V}N_{v}\times K_{\mathfrak{T}\backslash T}\subset V_{0}$ hence
$F_{T,x}(V_{0}^{\prime})\leq E_{x}(V_{0})$. To summarize, we have found a
compact $V_{1}^{\prime}\subset K_{T}$ and an open $V_{0}^{\prime}\subset
K_{T}$ such that $V_{1}^{\prime}\subset V\subset V_{0}^{\prime}$ and
\[
F_{T,x}(V_{0}^{\prime})-\varepsilon<F_{T,x}(V)<F_{T,x}(V_{1}^{\prime
})+\varepsilon
\]
which means that $F_{T,x}$ is regular.

Now we can prove the theorem. For any $f\in C(K_{T})$ we have
\begin{align*}
\int_{K_{T}}fdF_{T,x}  &  =\int_{K}\left(  f\circ\iota_{T}\right)  dE_{x}\\
&  =\left\langle x,\varphi_{T}\left(  f\right)  x\right\rangle \\
&  =\int_{K_{T}}fdE_{T,x}%
\end{align*}
and since both $E_{T,x}$ and $F_{T,x}$ are regular, we then know from Riesz's
representation theorem that $E_{T,x}=F_{T,x}$. In other words
\[
\left\langle x,E_{T}(V)x\right\rangle =\left\langle x,F_{T}(V)x\right\rangle
\]
for all $x\in H$ and all Borel $V\subset K_{T}$, hence by the polarization
identity $E_{T}=F_{T}$. $\blacksquare$

\bigskip

\noindent\textbf{Corollary 3.3. }\textit{Extend }$\varphi_{T}$\textit{ in
Theorem 3.2 to a unital }$\ast$\textit{-homomorphism }$\tilde{\varphi}%
_{T}:B_{\infty}(K_{T})\rightarrow L(H)$\textit{ defined by }%
\[
\tilde{\varphi}_{T}(f):=\int_{K_{T}}fdE_{T}%
\]
\textit{for all }$T\subset\mathfrak{T}$\textit{, and write }$\tilde{\varphi
}:=\tilde{\varphi}_{\mathfrak{T}}$\textit{ which is therefore the extension of
}$\varphi$\textit{ given by }$E$\textit{. Set }$\tilde{\psi}_{T}:B_{\infty
}(K_{T})\rightarrow B_{\infty}(K):f\mapsto f\circ\iota_{T}$\textit{. Then
}$\tilde{\varphi}_{T}=\tilde{\varphi}\circ\tilde{\psi}_{T}$\textit{ for all
non-empty }$T\subset\mathfrak{T}$\textit{.}

\bigskip

\noindent\textit{Proof.} Note that $\tilde{\psi}_{T}$ is well defined, since
$\iota_{T}$ is continuous. For any $x\in H$ we have using the notation of the
previous proof that
\begin{align*}
\left\langle x,\tilde{\varphi}_{T}(f)x\right\rangle  &  =\int_{K_{T}}%
fdE_{T,x}\\
&  =\int_{K_{T}}fd\left(  E_{x}\circ\iota_{T}^{-1}\right) \\
&  =\int_{K}\left(  f\circ\iota_{T}\right)  dE_{x}\\
&  =\left\langle x,\tilde{\varphi}\circ\tilde{\psi}_{T}(f)x\right\rangle
\end{align*}
for all $f\in B_{\infty}(K_{T})$. Hence by the polarization identity
$\tilde{\varphi}_{T}=\tilde{\varphi}\circ\tilde{\psi}_{T}$. $\blacksquare$

\bigskip

We also have the following simple proposition regarding spectral resolutions:

\bigskip

\noindent\textbf{Proposition 3.4.} \textit{Let }$K$\textit{ be a compact
Hausdorff space, }$H$\textit{ a Hilbert space, }$\varphi:C(K)\rightarrow
L(H)$\textit{ a unital }$\ast$\textit{-homomorphism and }$E$\textit{ its
spectral resolution. For any unitary }$U:H\rightarrow H$\textit{ set }%
$\varphi^{\prime}:=U^{\ast}\varphi(\cdot)U$\textit{ and let }$E^{\prime}%
$\textit{ be its spectral resolution. Then }$E^{\prime}=U^{\ast}E(\cdot
)U$\textit{. Furthermore, for the situation in Theorem 3.2 and Corollary 3.3,
and with }$E_{T}^{\prime}$\textit{ the spectral resolution of }$\varphi
_{T}^{\prime}:=\varphi^{\prime}\circ\psi_{T}$\textit{, we have }$E_{T}%
^{\prime}=U^{\ast}E_{T}(\cdot)U$\textit{ and hence }$\tilde{\varphi}%
_{T}^{\prime}=U^{\ast}\tilde{\varphi}_{T}(\cdot)U$\textit{ with }%
$\tilde{\varphi}_{T}^{\prime}$\textit{ the extension of }$\varphi_{T}^{\prime
}$\textit{ to }$B_{\infty}\left(  K_{T}\right)  $\textit{ given by }%
$E_{T}^{\prime}$\textit{.}

\bigskip

\noindent\textit{Proof.} We use a similar argument as in Theorem 3.2's proof.
Let $F:=U^{\ast}E(\cdot)U$ then $F_{x,x}:=\left\langle x,F(\cdot
)x\right\rangle =E_{Ux,Ux}$ is a regular positive Borel measure on $K$ for all
$x\in H$ by the properties of $E$. Furthermore, for every $f\in C(K)$ and
every $x\in H$ we have $\int_{K}fdE_{x,x}^{\prime}=\left\langle x,\varphi
^{\prime}(f)x\right\rangle =\left\langle Ux,\varphi(f)Ux\right\rangle
=\int_{K}fdF_{x,x}$ but $E_{x,x}^{\prime}$ is also a regular positive Borel
measure on $K$ by definition, hence by Riesz's representation theorem and the
polarization identity $E^{\prime}=F$.

Now, for the situation in Theorem 3.2 and with $\tilde{\varphi}^{\prime
}:=\tilde{\varphi}_{\mathfrak{T}}^{\prime}$, consider any Borel $V\subset
K_{T}$, then $E_{T}^{\prime}(V)=\tilde{\varphi}_{T}^{\prime}\left(  \chi
_{V}\right)  =\tilde{\varphi}^{\prime}\circ\tilde{\psi}_{T}\left(  \chi
_{V}\right)  =\tilde{\varphi}^{\prime}\left(  \chi_{\iota_{T}^{-1}(V)}\right)
=E^{\prime}\left(  \iota_{T}^{-1}(V)\right)  =U^{\ast}E\left(  \iota_{T}%
^{-1}(V)\right)  U=U^{\ast}E_{T}(V)U$ by Corollary 3.3 and Theorem 3.2.

For any $f\in B_{\infty}\left(  K_{T}\right)  $ and $x\in H$ it now follows
(using the notation of Theorem 3.2's proof) that $\left\langle x,\tilde
{\varphi}_{T}^{\prime}(f)x\right\rangle =\int_{K_{T}}fdE_{T,x}^{\prime}%
=\int_{K}fdE_{T,Ux}=\left\langle Ux,\tilde{\varphi}_{T}(f)Ux\right\rangle
=\left\langle x,U^{\ast}\tilde{\varphi}_{T}(f)Ux\right\rangle $. $\blacksquare$

\bigskip

These results will be used in the next section.

The remainder of this section shows how the imbedding $\psi_{T}:C\left(
K_{T}\right)  \rightarrow C(K)$ that appears in Theorem 3.2, or to be more
precise, the extended embedding $\tilde{\psi}_{T}$ in Corollary 3.3, can be
carried over to the pure representations of $C\left(  K_{T}\right)  $ and
$C(K)$. This rounds off the discussion in this section, but is less important
for the rest of the paper. Note that by the term \textit{imbedding} we mean an
injective $\ast$-homomorphism from one $C^{\ast}$-algebra to another. By
standard theory of $C^{\ast}$-algebras such an imbedding is automatically norm
preserving, and its image a $C^{\ast}$-algebra.

Let $\mathfrak{T}$, $K_{t}$, $K_{T}$, $K$ and $\psi_{T}$ be as in Theorem 3.2,
let $\left(  H_{T},\theta_{T}\right)  $ be a pure representation of $C\left(
K_{T}\right)  $ for every non-empty $T\subset\mathfrak{T}$ and set
$(H,\pi):=\left(  H_{\mathfrak{T}},\theta_{\mathfrak{T}}\right)  $. Keep in
mind that $\tilde{\theta}_{T}\left(  B_{\infty}\left(  K_{T}\right)  \right)
\subset L\left(  H_{T}\right)  $ is a $C^{\ast}$-algebra, since $B_{\infty
}\left(  K_{T}\right)  $ is a $C^{\ast}$-algebra and $\tilde{\theta}_{T}$ is a
$\ast$-homomorphism. We simply want to show that there is an imbedding
$\eta_{T}:\tilde{\theta}_{T}\left(  B_{\infty}\left(  K_{T}\right)  \right)
\rightarrow L(H)$ such that $\eta_{T}\circ\tilde{\theta}_{T}=\tilde{\pi}%
\circ\tilde{\psi}_{T}$, i.e. we have a commutative diagram, which means that
the imbedding $\tilde{\psi}_{T}:B_{\infty}\left(  K_{T}\right)  \rightarrow
B_{\infty}(K)$ has been carried over to $\tilde{\theta}_{T}\left(  B_{\infty
}\left(  K_{T}\right)  \right)  \rightarrow L(H)$ in a consistent way.

To do this it will be notationally convenient to write pure representations in
a slightly more concrete form. For any compact Hausdorff space $K$ it is
easily seen that the GNS representation of the pure state $\omega_{x}$
previously used in constructing a pure representation $(H,\pi)$ of $C(K)$ can
be taken to be $(H_{x},\pi_{x},\Omega_{x})=\left(  \mathbb{C},\omega
_{x},1\right)  $ where here we view $\omega_{x}(f)$ as an element of
$L(\mathbb{C})$ by $\mathbb{C}\rightarrow\mathbb{C}:z\mapsto\omega_{x}(f)z$
for all $f\in C(K)$. This pure representation is then given by%
\begin{equation}
H:=\bigoplus_{x\in K}\mathbb{C} \tag{3.3}%
\end{equation}
and
\begin{equation}
\pi:=\bigoplus_{x\in K}\omega_{x} \tag{3.4}%
\end{equation}
so $\pi(f)=\bigoplus_{x\in K}f(x)$ for all $f\in C(K)$ where $\bigoplus_{x\in
K}a_{x}$, with any bounded $K\ni x\mapsto a_{x}\in\mathbb{C}$, denotes an
element of $L(H)$ defined by $\left(  \bigoplus_{x\in K}a_{x}\right)  \left(
v_{x}\right)  _{x\in K}:=\left(  a_{x}v_{x}\right)  _{x\in K}$ for $\left(
v_{x}\right)  _{x\in K}\in H$. When we use the form (3.3) and (3.4) we will
say that $\pi$ is in \textit{diagonal form}. Since the GNS representation of
$\omega_{x}$ can always be written in the form $\left(  \mathbb{C},\omega
_{x},1\right)  $, we know that a pure representation can always be written in
diagonal form.

\bigskip

\noindent\textbf{Proposition 3.5.} \textit{Let }$K$\textit{ be a compact
Hausdorff space and }$(H,\pi)$\textit{ a pure representation of }%
$C(K)$\textit{ in diagonal form. Then }%
\[
\tilde{\pi}(f)=\bigoplus_{x\in K}f(x)
\]
\textit{for all }$f\in B_{\infty}(K)$\textit{.}

\bigskip

\noindent\textit{Proof.} Let $E$ be the spectral resolution of $\pi$ and
define $\Omega_{x}\in H$ by $\Omega_{x}:=\left(  \Omega_{x,y}\right)  _{y\in
K}$ where $\Omega_{x,y}=1$ for $y=x$ and $\Omega_{x,y}=0$ otherwise. Hence
$\Omega_{x}$, $x\in K$, is the total orthonormal set in $H$ that we used
previously (but now in $\pi$ 's diagonal form). Setting $E_{x,y}:=\left\langle
\Omega_{x},E(\cdot)\Omega_{y}\right\rangle $ for all $x,y\in K$ it follows
from Theorem 3.1 that
\[
E_{x,x}(V)=\left\{
\begin{array}
[c]{c}%
1\text{ if }x\in V\\
0\text{ if }x\notin V
\end{array}
\right.
\]
for all Borel $V\subset K$, while $E_{x,y}=0$ for $x\neq y$. Hence
\[
\left\langle \Omega_{x},\tilde{\pi}\left(  f\right)  \Omega_{y}\right\rangle
=\int_{K}fdE_{x,y}=f(x)\left\langle \Omega_{x},\Omega_{y}\right\rangle
=\left\langle \Omega_{x},\left[  \bigoplus_{z\in K}f(z)\right]  \Omega
_{y}\right\rangle
\]
and since $\Omega_{x}$, $x\in K$, is a total orthonormal set in $H$, the
result follows. $\blacksquare$

\bigskip

\noindent\textbf{Corollary 3.6.} \textit{Let }$K$\textit{ be a compact
Hausdorff space and }$(H,\pi)$\textit{ a pure representation of }%
$C(K)$\textit{, then }$\tilde{\pi}:B_{\infty}(K)\rightarrow L(H)$\textit{ is injective.}

\bigskip

\noindent\textit{Proof.} Without loss we can put $\pi$ in diagonal form, hence
$\tilde{\pi}$ is given by Proposition 3.5. Now for any bounded $K\ni x\mapsto
a_{x}\in\mathbb{C}$ and $K\ni x\mapsto b_{x}\in\mathbb{C}$ with $\bigoplus
_{x\in K}a_{x}=\bigoplus_{x\in K}b_{x}$ as elements of $L(H)$, we have
\[
a_{x}=\left\langle \Omega_{x},\left(  \bigoplus_{y\in K}a_{y}\right)
\Omega_{x}\right\rangle =\left\langle \Omega_{x},\left(  \bigoplus_{y\in
K}b_{y}\right)  \Omega_{x}\right\rangle =b_{x}%
\]
for every $x\in K$, with $\Omega_{x}$ as in Proposition 3.5's proof. In
particular, if $\tilde{\pi}(f)=\tilde{\pi}(g)$ for $f,g\in B_{\infty}(K)$,
then $f=g$. $\blacksquare$

\bigskip

\noindent\textbf{Corollary 3.7.}\textit{ For the situation in Theorem 3.2, but
with }$\varphi=\pi$\textit{ a pure representation of }$C(K)$\textit{, it
follows that }$\tilde{\pi}_{T}$\textit{ is injective, and with }$\pi$\textit{
in diagonal form it is given by }%
\begin{equation}
\tilde{\pi}_{T}(f)=\bigoplus_{x\in K}f\left(  \iota_{T}(x)\right)  \tag{3.5}%
\end{equation}
\textit{for all }$f\in B_{\infty}\left(  K_{T}\right)  $\textit{.}

\bigskip

\noindent\textit{Proof.} By Corollary 3.3 we have $\tilde{\pi}_{T}%
(f)=\tilde{\pi}\left(  f\circ\iota_{T}\right)  $, and since $\tilde{\pi}$ is
injective by Corollary 3.6 while $\iota_{T}$ is surjective, it follows that
$\tilde{\pi}_{T}$ is injective. With $\pi$ in diagonal form, (3.5) follows
immediately from Corollary 3.3 and Proposition 3.5. $\blacksquare$

\bigskip

\noindent\textbf{Theorem 3.8.}\textit{ For the situation in Theorem 3.2, let
}$\left(  H_{T},\theta_{T}\right)  $\textit{ be a pure representation of
}$C(K_{T})$\textit{, and set }$\pi:=\theta_{\mathfrak{T}}$\textit{ and }%
$\pi_{T}:=\pi\circ\psi_{T}$\textit{ for every non-empty }$T\subset
\mathfrak{T}$\textit{. Then for every such }$T$\textit{ there is a unique
function }%
\[
\eta_{T}:\tilde{\theta}_{T}\left(  B_{\infty}\left(  K_{T}\right)  \right)
\rightarrow L(H)
\]
\textit{such that }$\eta_{T}\circ\tilde{\theta}_{T}=\tilde{\pi}_{T}$\textit{.
Furthermore, }$\eta_{T}$\textit{ is an injective norm preserving unital }%
$\ast$\textit{-homomorphism. With }$\pi$\textit{ and }$\theta_{T}$\textit{ in
diagonal form, it is given by}%
\begin{equation}
\eta_{T}\left(  \bigoplus_{x\in K_{T}}f(x)\right)  =\bigoplus_{x\in K}f\left(
\iota_{T}(x)\right)  \tag{3.6}%
\end{equation}
\textit{for all }$f\in B_{\infty}\left(  K_{T}\right)  $\textit{.}

\bigskip

\noindent\textit{Proof.} The existence and uniqueness of $\eta_{T}$ follows
from the injectivity of $\tilde{\theta}_{T}$ given by Corollary 3.6. It is a
$\ast$-homomorphism, since $\tilde{\theta}_{T}$ and $\tilde{\pi}_{T}$ are, and
it is injective, since $\tilde{\pi}_{T}$ is injective according to Corollary
3.7. Hence $\eta_{T}$ is norm preserving. It is unital since $\tilde{\theta
}_{T}$ and $\tilde{\pi}_{T}$ are. In diagonal form $\tilde{\theta}_{T}$ is
given by Proposition 3.5, hence (3.6) follows directly from (3.5).
$\blacksquare$

\bigskip

\noindent\textbf{Corollary 3.9.}\textit{ For the situation in Theorem 3.8, and
with }$E_{T}$\textit{ and }$F_{T}$\textit{ the spectral resolutions of }%
$\pi_{T}$\textit{ and }$\theta_{T}$\textit{ respectively, we have }$\eta
_{T}\circ F_{T}=E_{T}$\textit{.}

\bigskip

\noindent\textit{Proof.} For any Borel $V\subset K_{T}$ we have $\eta
_{T}\left(  F_{T}(V)\right)  =\eta_{T}\left(  \tilde{\theta}_{T}\ (\chi
_{V})\right)  =\tilde{\pi}_{T}\left(  \chi_{V}\right)  =E_{T}(V)$.
$\blacksquare$

\section{Evolution integrals}

Now we apply the ideas of the previous section to find an analogue of path
integrals for abstract dynamical systems. Fix an arbitrary set $\mathfrak{T}$.
We will allow $T$ to be any subset of $\mathfrak{T}$. We will view
$\mathfrak{T}$ as the set of all points in ``time'' (corresponding to
$\mathbb{R}$ in usual quantum mechanics), and the $T$ 's as ``time
intervals''. Let $A$ be a $W^{\ast}$-algebra as in Section 2, and set
\[
X_{T}:=\overline{\text{Aut}(A)}^{T}%
\]
which is the evolution space over $T$. Write $X:=X_{\mathfrak{T}}$. Our goal
is to do integrals over $X_{T}$ to represent dynamics on a ``higher level'' as
discussed in the introduction, the higher level being a Hilbert space obtained
from a pure representation. However, we would like to use the same Hilbert
space for different $T$, since then we can interpret the integrals for
different $T$ 's to represent the dynamics of the same system but over
different ``time intervals''. Therefore we will imbed $C(X_{T})$ canonically
into $C(X)$ and then consider a pure representation of $C(X)$. To do this let
$\iota_{T}$ and $\psi_{T}$ be defined as in Theorem 3.2 in terms of $K=X$ and
$K_{T}=X_{T}$. Then $\psi_{T}$ is a well defined injective norm preserving
unital $\ast$-homomorphism, which can be viewed as a canonical imbedding of
$C(X_{T})$ into $C(X)$. Let $(H,\pi)$ be a pure representation of $C(X)$,
which makes $H$ independent of $T$, and then consider the spectral resolution
$E_{T}:=E_{\pi_{T}}$ of the injective unital $\ast$-homomorphism
\[
\pi_{T}:=\pi\circ\psi_{T}:C(X_{T})\rightarrow L(H)\text{.}%
\]
This spectral measure $E_{T}$ can be viewed as being natural, since $\pi$ and
$\psi_{T}$ are both canonical (also see Section 3), and allows us to do
integrals over the evolution space $X_{T}$, namely
\begin{equation}
\tilde{\pi}_{T}(f)=\int_{X_{T}}fdE_{T} \tag{4.1}%
\end{equation}
is defined for all $f\in B_{\infty}(X_{T})$. We will call integrals of the
form (4.1) \textit{evolution integrals}.

Theorem 3.8 tells us that it would in fact be mathematically equivalent to
work on the various Hilbert spaces $H_{T}$, $T\subset\mathfrak{T}$, however we
will express everything in terms of $H$, since this is a simpler point of view
as far as the dynamical system on the higher level is concerned.

The arguments in Section 3 that $E:=E_{\mathfrak{T}}$ attaches the same weight
to all the points of $K=X$ can be interpreted as each evolution over
$\mathfrak{T}$ having the same weight. Via Theorem 3.2 we can then also say
that $E_{T}$ attaches the same weight to each of the evolutions over $T$,
namely to each point of $X_{T}$. This is analogous to path integrals, as
explained at the beginning of Section 3, and hence is exactly the type of
structure that we intuitively want.

Our interpretation of $H$ in a pure representation of $C(K)$ in Section 3
gives us a nice picture in the case where $K=X$, namely the vectors $\left\{
\Omega_{\alpha}^{\prime}\right\}  _{\alpha\in X}$ defined as in the case of
$K$, represent the evolutions of $A$ over the entire $\mathfrak{T}$ as a total
orthonormal set in $H$. In a similar way an evolution $\beta\in X_{T}$
corresponds to the set of evolutions in $X$ projected onto $\beta$ by
$\iota_{T}$, and hence to the set of $\Omega_{\alpha}^{\prime}$ 's with
$\iota_{T}(\alpha)=\beta$.

The basic idea for getting dynamics on $H$, is to consider a unitary $u_{T}\in
B_{\infty}(X_{T})$, and then set $U_{T}=\tilde{\pi}_{T}(u_{T})$ which is a
unitary operator on $H$, since $\tilde{\pi}_{T}$ is a unital $\ast
$-homomorphism. We will interpret $U_{T}$ as representing dynamics on $H$ over
the set $T$, and will discuss this in more detail below, and in the next section.

Note that since $B_{\infty}(X_{T})$ is abelian and $\tilde{\pi}_{T}$ is a
homomorphism, all $U_{T}$ 's obtained in this way will commute with each
other. This is where the fact that a pure representation of $C(K)$ is not
unique, as mentioned in Section 3, comes into play. To obtain unitaries on $H$
which do not commute with these $U_{T}$ 's, we can use a unitary
transformation of $\pi$ to get another pure representation of $C(K)$, namely
$\pi^{\prime}:=U^{\ast}\pi(\cdot)U$ with $U$ a unitary operator on $H$, and
then replace $E_{T}$ by the spectral resolution $E_{T}^{\prime}$ of $\pi
_{T}^{\prime}:=\pi^{\prime}\circ\psi_{T}$. By Proposition 3.4 we get
$U_{T}^{\prime}:=\int_{X_{T}}u_{T}dE_{T}^{\prime}=U^{\ast}U_{T}U$ instead of
$U_{T}$. In this sense evolution integrals differ from path integrals. Instead
of having one $\mathbb{R}^{+}\cup\{\infty\}$-valued measure, we have many
projection valued measures.

Although we will not discuss detailed examples and applications in this paper,
in this paragraph we give a brief description of how simple examples can be
obtained, before we resume with the theory. As mentioned above we are
interested in unitary operators on $H$ given by $U_{T}=\tilde{\pi}_{T}\left(
u_{T}\right)  $ with $u_{T}\in B_{\infty}\left(  X_{T}\right)  $ unitary. A
simple case of this would be $u_{T}=e^{iS_{T}}$ where $S_{T}:X_{T}%
\rightarrow\mathbb{R}$ is continuous. Hence we look at a simple class of
examples of such an $S_{T}$. Let $T$ be a finite subset of $\mathfrak{T}$ and
let $f_{t}\in L(A)_{\ast}$ for every $t\in T$. Let $g_{t}:\mathbb{C}%
\rightarrow\mathbb{R}$ be continuous for every $t\in T$, for example
$g_{t}=\left|  \cdot\right|  $ or $g_{t}=\left|  \cdot\right|  ^{2}$. Fix any
$\left(  \tau_{t}\right)  _{t\in T}\in X_{T}$. For $f\in L(A)_{\ast}$ and
$\alpha\in\overline{\text{Aut}(A)}$ we view $\alpha$ as a linear functional on
$L(A)_{\ast}$ and denote its value at $f$ by $f(\alpha)$. Now set
$S_{T}(\alpha):=\sum_{t\in T}g_{t}\left(  f_{t}(\alpha_{t}-\tau_{t})\right)  $
for all $\alpha=\left(  \alpha_{t}\right)  _{t\in T}\in X_{T}$. Then $S_{T}$
is continuous, since $T$ is finite and we are using the product topology on
$X_{T}$. This example gives some indication that we would have to be careful
when attempting to construct examples for the case where $T$ is not finite.

Next we refine our idea for obtaining dynamics on $H$, by taking
$\mathfrak{T}$ to be a measure space $(\mathfrak{T},\Sigma,\mu)$ with $\Sigma$
a $\sigma$-algebra in $\mathfrak{T}$ and $\mu$ a usual positive measure on
$\Sigma$. In the following section this will allow us to clarify the analogy
with path integrals. Let $\mathcal{U}(\mathfrak{A})$ denote the set of all
unitary elements of any unital $C^{\ast}$-algebra $\mathfrak{A}$.

\bigskip

\noindent\textbf{Definition 4.1.} Let $\left(  \mathfrak{T},\Sigma,\mu\right)
$ be a measure space, and $\Sigma_{0}\subset\Sigma$ a set such that $T_{1}\cup
T_{2}\in\Sigma_{0}$ when $T_{1},T_{2}\in\Sigma_{0}$. Now consider a function%
\[
u:\Sigma_{0}\rightarrow\bigcup_{T\in\Sigma_{0}}B_{\infty}(X_{T}):T\mapsto
u_{T}%
\]
such that
\[
u_{T}\in\mathcal{U}\left(  B_{\infty}(X_{T})\right)
\]
for all $T\in\Sigma_{0}$\textit{,}%
\begin{equation}
u_{T_{1}\cup T_{2}}\left(  \iota_{T_{1}\cup T_{2}}(\alpha)\right)  =u_{T_{1}%
}\left(  \iota_{T_{1}}(\alpha)\right)  u_{T_{2}}\left(  \iota_{T_{2}}%
(\alpha)\right)  \tag{4.2}%
\end{equation}
for all $\alpha\in X$ and all $T_{1},T_{2}\in\Sigma_{0}$ with $\mu(T_{1}\cap
T_{2})=0$, and
\begin{equation}
u_{T}=1 \tag{4.3}%
\end{equation}
for all $T\in\Sigma_{0}$ with $\mu(T)=0$. Such a $u$, or $\left(  u,\Sigma
_{0},\mu\right)  $ to be more complete, will be called an \textit{action
weight} for $\left(  A,\mathfrak{T}\right)  $. If $u_{T}\in\mathcal{U}\left(
C(X_{T})\right)  $ for all $T\in\Sigma_{0}$, we will call $u$ a
\textit{continuous action weight}.

\bigskip

The word ``action'' in the term action weight is borrowed from the classical
action which appears in usual path integrals in quantum mechanics. Note that
when using an action weight, we no longer allow all subsets $T\subset
\mathfrak{T}$, but only $T\in\Sigma_{0}$. A typical situation might be where
$\mathfrak{T}$ is a locally compact Hausdorff group, $\Sigma$ its Borel
$\sigma$-algebra, $\mu$ its Haar-measure, and $\Sigma_{0}$ the Borel sets with
finite measure.

So let $u$ be an action weight as in Definition 4.1, and define
\begin{equation}
U_{T}=\int_{X_{T}}u_{T}dE_{T} \tag{4.4}%
\end{equation}
for all $T\in\Sigma_{0}$ with the convention that $U_{\varnothing}=1$ if
$\varnothing\in\Sigma_{0}$. Note that $U_{T}$ is defined in terms of a fixed
pure representation $\pi$ of $C(X)$, so all the $U_{T}$ 's commute with on
another as discussed earlier. Although $u$ corresponds to the classical level
in path integrals, and $U$ to the quantum level, one should keep in mind that
$A$ need not be abelian, hence we in general do not view $u$ as coming from
``classical'' dynamics, unless $A$ is abelian. Next we show that the dynamics
on $H$ given by the $U_{T}$ 's have familiar grouplike properties:

\bigskip

\noindent\textbf{Proposition 4.2.} \textit{Let }$U_{T}$\textit{ be defined as
in (4.4). Then we have}
\[
U_{T_{1}}U_{T_{2}}=U_{T_{1}\cup T_{2}}%
\]
\textit{for all }$T_{1},T_{2}\in\Sigma_{0}$\textit{ with }$\mu\left(
T_{1}\cap T_{2}\right)  =0$\textit{, and}
\[
U_{T}=1
\]
\textit{for all }$T\in\Sigma_{0}$\textit{ with }$\mu(T)=0$\textit{.}

\bigskip

\noindent\textit{Proof.} \noindent If any of $T_{1}$, $T_{2}$ or $T$ are
empty, the result is trivial, so assume they are not empty. Using Corollary
3.3 and its notation, we have
\begin{align*}
U_{T_{1}}U_{T_{2}}  &  =\tilde{\pi}_{T_{1}}(u_{T_{1}})\tilde{\pi}_{T_{2}%
}(u_{T_{2}})\\
&  =\tilde{\pi}\left(  \tilde{\psi}_{T_{1}}(u_{T_{1}})\right)  \tilde{\pi
}\left(  \tilde{\psi}_{T_{2}}(u_{T_{2}})\right) \\
&  =\tilde{\pi}\left(  \tilde{\psi}_{T_{1}}(u_{T_{1}})\tilde{\psi}_{T_{2}%
}(u_{T_{2}})\right) \\
&  =\tilde{\pi}\left(  \tilde{\psi}_{T_{1}\cup T_{2}}(u_{T_{1}\cup T_{2}%
})\right) \\
&  =U_{T_{1}\cup T_{2}}%
\end{align*}
by (4.2). From (4.3) on the other hand, we immediately have $U_{T}=\tilde{\pi
}_{T}(u_{T})=\tilde{\pi}_{T}(1)=1$. $\blacksquare$

\bigskip

These are the properties that one would have in quantum mechanics coming from
path integrals over closed intervals, say $T_{1}=[t_{0},t_{1}]$ and
$T_{2}=[t_{1},t_{2}]$, with $\mathfrak{T}=\mathbb{R}$, $\mu$ the Lebesgue
measure on $\mathbb{R}$, and $\Sigma_{0}$ for example being the sets with
finite Lebesgue measure. There one would interpret these properties as the
group structure of the quantum dynamics, with inverses of the unitaries
providing the group inverse. Since all the $U_{T}$ 's commute, this grouplike
structure is inherently abelian.

\bigskip

\noindent\textit{Remarks.}

Any unital $\ast$-homomorphism $\varphi:C(X)\rightarrow L(H)$, with $H$ any
Hilbert space, in principle gives us another definition of evolution integrals
if we replace $\pi$ by $\varphi$, but $\pi$ seems most canonical, and also
gives a very simple interpretation of $H$ as discussed above. Note that
Proposition 4.2 still holds if we replace $\pi$ by such a $\varphi$. One other
choice besides $\pi$ that would be canonical, is the universal representation
of $C(X)$.

A different approach we could have followed in setting up our framework, is to
begin with a Hilbert space $H_{0}$ instead of $A$, and then consider the set
$\mathcal{U}(H_{0})$ of unitary operators $H_{0}\rightarrow H_{0}$ instead of
Aut$(A)$. Since $\mathcal{U}(H_{0})\subset L(H_{0})$ with $L(H_{0})$ a von
Neumann algebra which therefore has a unique predual, we could look at the
closure $\overline{\mathcal{U}(H_{0})}$ in the resulting weak* topology on
$L(H_{0})$. Then replace $\overline{\text{Aut}(A)}^{T}$ in the framework we
set up above with the compact Hausdorff space $\overline{\mathcal{U}(H_{0}%
)}^{T}$, the elements of which we would interpret as evolutions in $H_{0}$.
However, to start with $A$ instead of $H_{0}$ seems more natural simply
because the algebraic formulation is a more natural way to formulate a
dynamical system, with a Hilbert space being a specific way of representing
such a system. This is especially clear if we start with a classical system on
the lower level. This raises the question if the dynamical system on the
higher level, namely the Hilbert space $H$ and the dynamics on it, is in some
natural way a representation of an algebraic formulation of the same dynamical system.

\section{The analogy with path integrals}

We now make the analogy between evolution integrals and path integrals more
explicit. We will continue using the notation from the previous section. In
particular, $\Sigma$ is still a $\sigma$-algebra in $\mathfrak{T}$. For
$T\in\Sigma$, set $\Sigma|_{T}:=\left\{  V\cap T:V\in\Sigma\right\}  $ which
is a $\sigma$-algebra in $T$. Denote the vector space of bounded $\Sigma|_{T}%
$-measurable functions $T\rightarrow\mathbb{C}$ with the sup-norm by
$B_{\infty}(\Sigma|_{T})$. Although this space is a $C^{\ast}$-algebra, we
will only use its normed space structure.

\bigskip

\noindent\textbf{Definition 5.1.} Let\textit{ }$\Sigma_{0}\subset\Sigma$ be
the sets with finite $\mu$-measure. Consider a mapping $\mathcal{L}%
:T\rightarrow\mathcal{L}_{T}$ on $\Sigma_{0}$ such that $\mathcal{L}_{T}%
:X_{T}\rightarrow B_{\infty}\left(  \Sigma|_{T}\right)  :\alpha\mapsto
\mathcal{L}_{T,\alpha}$ is continuous, $\mathcal{L}_{T,\alpha}$ is real-valued
and $\mathcal{L}_{T,\alpha}|_{T^{\prime}}=\mathcal{L}_{T^{\prime}%
,\alpha|_{T^{\prime}}}$ for all $T,T^{\prime}\in\Sigma_{0}$ with $T^{\prime
}\subset T$. We will call an $\mathcal{L}$ with these properties a
\textit{Lagrangian}. Define $S_{T}:X_{T}\rightarrow\mathbb{R}$ by\textit{ }%
\[
S_{T}(\alpha):=\int_{T}\mathcal{L}_{T,\alpha}d\mu
\]
for all $\alpha\in X_{T}$ and all $T\in\Sigma_{0}$\textit{.} The mapping
$S:T\mapsto S_{T}$ defined on $\Sigma_{0}$ will be called the \textit{action}
of $\mathcal{L}$\textit{.}

\bigskip

The terminology in this definition is of course borrowed from classical
mechanics. The simple example given in Section 4 is such an action.

\bigskip

\noindent\textbf{Proposition 5.2.}\textit{ The function }$S_{T}$\textit{ in
Definition 5.1 is continuous, hence }$S$\textit{ is a function }$\Sigma
_{0}\rightarrow\bigcup_{T\in\Sigma_{0}}C(X_{T})$\textit{ with }$S_{T}\in
C(X_{T})$\textit{ for every }$T\in\Sigma_{0}$\textit{. Setting }%
$u_{T}:=e^{iS_{T}}$\textit{ for all }$T\in\Sigma_{0}$\textit{ makes the
function }$u$\textit{ given by }$\Sigma_{0}\ni T\mapsto u_{T}$\textit{ a
continuous action weight.}

\bigskip

\noindent\textit{Proof.} For any $\alpha,\beta\in X_{T}$ we have
\begin{align*}
\left|  S_{T}(\alpha)-S_{T}(\beta)\right|   &  =\left|  \int_{T}\left(
\mathcal{L}_{T,\alpha}-\mathcal{L}_{T,\beta}\right)  d\mu\right| \\
&  \leq\left\|  \mathcal{L}_{T,\alpha}-\mathcal{L}_{T,\beta}\right\|  \mu(T)
\end{align*}
and since $\mu(T)<\infty$ while $\mathcal{L}_{T}$ is continuous by assumption,
we know that for every $\varepsilon>0$ there is a neighbourhood $N$ of
$\alpha$ in $X$ such that
\[
\left|  S_{T}(\alpha)-S_{T}(\beta)\right|  <\varepsilon
\]
for all $\beta\in N$. Hence $S_{T}$ is continuous. Since $S_{T}$ is continuous
and real-valued, we have $u_{T}\in\mathcal{U}\left(  C(X_{T})\right)  $. For
$T_{1},T_{2}\in\Sigma_{0}$ with $\mu(T_{1}\cap T_{2})=0$ we have for all
$\alpha\in X$ that
\begin{align*}
S_{T_{1}\cup T_{2}}\left(  \iota_{T_{1}\cup T_{2}}(\alpha)\right)   &
=\int_{T_{1}\cup T_{2}}\mathcal{L}_{T_{1}\cup T_{2},\iota_{T_{1}\cup T_{2}%
}(\alpha)}d\mu\\
&  =\int_{T_{1}}\mathcal{L}_{T_{1},\iota_{T_{1}}(\alpha)}d\mu+\int_{T_{2}%
}\mathcal{L}_{T_{2},\iota_{T_{2}}(\alpha)}d\mu\\
&  =S_{T_{1}}\left(  \iota_{T_{1}}(\alpha)\right)  +S_{T_{2}}\left(
\iota_{T_{2}}(\alpha)\right)
\end{align*}
from which (4.2) follows. For $T\in\Sigma_{0}$ with $\mu(T)=0$ we immediately
have $S_{T}=0$, and (4.3) follows. $\blacksquare$

\bigskip

The evolution integral (4.4) takes the form
\[
U_{T}=\int_{X_{T}}e^{iS_{T}}dE_{T}%
\]
for all $T\in\Sigma_{0}$ in the case of the action weight given by Proposition
5.2, which makes the analogy with path integrals in quantum mechanics
particularly clear, although a path integral gives an amplitude and therefore
more properly corresponds to $\left\langle x,U_{T}y\right\rangle $ with
$x,y\in H$.

\section{Concluding remarks}

There are a number of aspects of evolution integrals that could merit further
investigation. For example, is it possible to use only continuous evolutions
$T\rightarrow\overline{\text{Aut}(A)}$ where we assume $T$ is a topological
space? Or to use Aut$(A)^{T}$ instead of $\overline{\text{Aut}(A)}^{T}$? Also,
if given a Hilbert space representation of an abstract dynamical system, is
there a way to decide whether or not its dynamics is given by evolution
integrals over some other system with certain properties, for example having
an abelian $W^{\ast}$-algebra? And can evolution integrals be linked more
closely with quantum physics, for example by trying to find the connection
between the Hilbert space $H$ used above and the usual quantum state space in
the case where $A$ is abelian and represents a classical physical system?

\bigskip

\textit{Acknowledgements. }I thank Joe Diestel, Johan Swart and Gusti van Zyl
for useful discussions concerning tensor products and preduals of Banach spaces.

\end{document}